\documentclass[sigconf]{acmart}
\renewcommand\footnotetextcopyrightpermission[1]{} 
\usepackage{tabularx}
\usepackage{sidecap}

\usepackage{nopageno}
\begin{document}
	\title{Understanding, Categorizing and Predicting Semantic Image-Text 
	Relations}
	
	\author{Christian Otto}
	\orcid{0000-0003-0226-3608}
	\affiliation{%
		\institution{\normalsize Leibniz Information Centre for Science and 
		Technology (TIB)}
		\city{Hannover}
		\country{Germany}
	}
	\author{Matthias Springstein}
	\orcid{0000-0002-6509-8534}
	\affiliation{%
		\institution{\normalsize Leibniz Information Centre for Science and 
		Technology (TIB)}
		\city{Hannover}
		\country{Germany}
	}	
	\author{Avishek Anand}
	\affiliation{%
		\institution{\normalsize L3S Research Center, Leibniz Universit\"at 
		Hannover}
		\city{Hannover}
		\country{Germany}
	}
	\author{Ralph Ewerth}
	\orcid{0000-0003-0918-6297}
	\affiliation{%
		\institution{\normalsize Leibniz Information Centre for Science and 
		Technology (TIB)}
		\institution{\normalsize L3S Research Center, Leibniz Universit\"at 
		Hannover}
		\city{Hannover}
		\country{Germany}
	}

\begin{abstract}
Two modalities are often used to convey information in a complementary and beneficial manner, e.g., in online news, videos, educational resources, or scientific publications. The automatic understanding of semantic correlations between text and associated images as well as their interplay has a great potential for enhanced multimodal web search and recommender systems. However, automatic understanding of multimodal information is still an unsolved research problem. Recent approaches such as image captioning focus on precisely describing visual content and translating it to text, but typically address neither semantic interpretations nor the specific role or purpose of an image-text constellation. In this paper, we go beyond previous work and investigate, inspired by research in visual communication, useful semantic image-text relations for multimodal information retrieval. We derive a categorization of eight semantic image-text classes (e.g., "illustration" or "anchorage") and show how they can systematically be characterized by a set of three metrics: cross-modal mutual information, semantic correlation, and the status relation of image and text. Furthermore, we present a deep learning system to predict these classes by utilizing multimodal embeddings. To obtain a sufficiently large amount of training data, we have automatically collected and augmented data from a variety of data sets and web resources, which enables future research on this topic. Experimental results on a demanding test set demonstrate the feasibility of the approach.

\end{abstract}

	\maketitle

	\renewcommand{\shortauthors}{}

	\section{Introduction}
\label{sec:introduction}
\begin{SCfigure}[1][!ht]
	\includegraphics[width=0.18\textwidth]{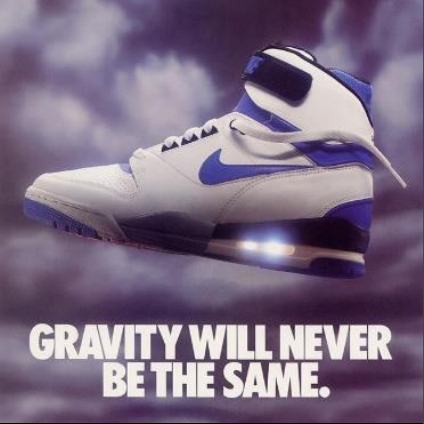}
	\hfill
	\caption{An example of a complex message portrayed by an image-text pair 
	elucidating the semantic gap between the textual information and the image 
	content. (Source:~\cite{hussain2017automatic})}
	\label{fig:divide}
\end{SCfigure}

The majority of textual web content is supplemented by multimedia information depicted in pictures, animations, audio, or videos -- for a good reason: it can help users to comprehend information more effectively and efficiently. In addition, some kind of information can be solely expressed by text and not by an image (e.g., a date like a person's birthday), or vice versa (e.g., the exact shape of a plant's leaf). Although multimodal information is omnipresent (for example, in web documents, videos, scientific papers, graphic novels), today's search engines and recommender systems do not exploit the full potential of multimodal information yet. When documents of different media types are retrieved to answer a user query, typically the results are displayed separately or sequentially, while semantic cross-modal relations are not exploited and remain hidden. One reason is that the automatic understanding and interpretation of (non-textual) visual or audio sources itself is difficult -- and it is even more difficult to model and understand the interplay of two different modalities (see Figure \ref{fig:divide}). Communication sciences and applied linguistics have been investigating the visual/verbal divide for many years (e.g., Barthes~\cite{barthes1977image}, Martinec and Salway~\cite{Martinec2005}, Bateman~\cite{bateman2017multimodality}). Although the semantic gap has been identified also as a fundamental problem in multimedia search nearly twenty years ago \cite{smeulders2000content}, insights and taxonomies from the field of visual communication science have been disregarded yet. 
However, we believe that insights from this field are very useful for multimodal retrieval research since they provide a new and differentiated perspective on image-text relations.
\vspace{0.2cm}

In this paper, we leverage taxonomies from visual communication research and derive a set of eight computable, semantic image-text relations for multimodal indexing and search. These image-text relations are systematically characterized by three metrics. Our contributions can be summarized as follows:

(1) \textit{Modeling a set of semantic image-text relations}: 
Based on previous work in communication sciences, we derive a categorization of \textit{distinctive} semantic image-text classes for multimodal analysis and search. To derive a systematic characterization, we build upon previous work in multimedia retrieval, where the metrics cross-modal mutual information and semantic correlation have been suggested to describe the information gap between image and text \cite{henning2017estimating}. In addition, we introduce a third metric, the status relation, and show that these three metrics allow us to systematically characterize the eight classes of image-text relations. 

(2) \textit{Training data augmentation}: 
Since there is no sufficiently large dataset set to train a deep learning system to predict the eight image-text classes, it is outlined how a comprehensive dataset can be automatically collected and augmented for this purpose.

(3) \textit{Automatic prediction of image-text classes}: Utilizing our new training dataset, we present a deep learning system to automatically classify these metrics. 
Two variations are realized and evaluated: (a) a "conventional" end-to-end approach for direct classification of an image-text class as well as (b) a "cascaded" architecture to estimate the different metrics separately and then infer the classes by combining the results. Experiments are conducted on a demanding, human-annotated testset. 



The remainder of the paper is organized as follows. Related work in the fields of communication sciences and information retrieval is discussed in Section 2. The derived categorization of image-text classes and their characterization by three dimensions are presented in Section 3. In Section 4, we propose a deep learning system to predict these metrics as well as resulting image-text classes and describe our approach for automatic data collection and augmentation. Experimental results are presented in Section 5, while Section 6 concludes the paper and outlines areas of future work.
	\section{Related Work}
\label{sec:relatedwork}

\subsection{Multimedia information retrieval}
Numerous publications in recent years deal with multimodal information in 
retrieval tasks. The general problem of reducing or bridging the semantic 
gap~\cite{smeulders2000content} between images and text is the main issue in 
cross-media retrieval~\cite{qi2018life, balaneshin2018deep, Mithun2018ACMMM, 
mithun2018learning, xu2018modal, joslyn2018cross}. Fan et al.~\cite{Fan2017} 
tackle this problem by modeling humans' visual and descriptive senses for an 
image through a multi-sensory fusion network. They argue to bridge the 
\textit{cognitive and semantic gap} by improving the comparability of 
heterogeneous media features and obtain good results for image-to-text and 
text-to-image retrieval. Liang et al.~\cite{Liang2016} propose a self-paced 
cross-modal subspace matching method by constructing a multimodal graph that 
preserves the intra-modality and inter-modality similarity. Another application 
is targeted by ~\cite{Mazloom2016}, who extract a set of engagement parameters 
to predict the popularity of social media posts. This can be leveraged by 
companies to understand their customers and evaluate marketing campaigns. 
While the confidence in predicting basic emotions like happiness or sadness can 
be improved by multimodal features~\cite{xu2017multisentinet}, even more 
complex semantic concepts like sarcasm~\cite{Schifanella2016} or 
metaphors~\cite{Shutova2016} can be predicted. This is enabled by evaluating 
the textual cues in the context of the image, providing a new level of semantic 
richness. The attention-based text embeddings introduced by Bahdanau et 
al.~\cite{bahdanau2014neural} analyze textual information under the 
consideration of previously generated image embedding and improve tasks like 
document classification~\cite{yang2016hierarchical} and image caption 
generation~\cite{xu2015show, johnson2016densecap, xie2014cross, lan2017fluency}.
Henning and Ewerth~\cite{henning2017estimating} propose two metrics to 
characterize image-text relations in a general manner: \textit{cross-modal 
mutual information} and \textit{semantic correlation}. They suggest an 
autoencoder with multimodal embeddings to learn these relations while 
minimizing the need for annotated training data.

A prerequisite to use heterogeneous modalities in machine learning approaches 
is the encoding in a joint feature space. The encoding might depend  on the 
type of modality to encode, the number of training samples available, the type 
of classification to perform and the desired interpretability of the 
models~\cite{baltruvsaitis2018multimodal}. One type of algorithms utilizes 
\textit{Multiple Kernel Learning}~\cite{bucak2014multiple, gonen2011multiple}. 
Application areas are multimodal affect recognition~\cite{poria2015deep, 
jaques2015multi}, event detection~\cite{yeh2012novel}, and Alzheimer's disease 
classification~\cite{liu2014multiple}. Deep neural networks can also be 
utilized to model multimodal embeddings. For instance, these systems can be 
used for the generation of image captions~\cite{karpathy2014deep}; Ramanishka 
et al.~\cite{Ramanishka2016} exploit audiovisual data and  metadata, i.e., a 
video's domain, to generate coherent video descriptions "in the wild", using 
convolutional neural networks (CNN, ResNet~\cite{he2016deep}) for encoding 
visual data. Alternative network architectures are 
GoogleNet~\cite{szegedy2017inception} or DenseNet~\cite{huang2016densely}.

\subsection{Communication sciences}
The interpretation of multimodal information and the "visual/verbal divide" 
have been investigated in the field of visual communication and applied 
linguistics for many years.

\begin{figure}[h!]
 	\centering
 	\includegraphics[width=0.4\textwidth]{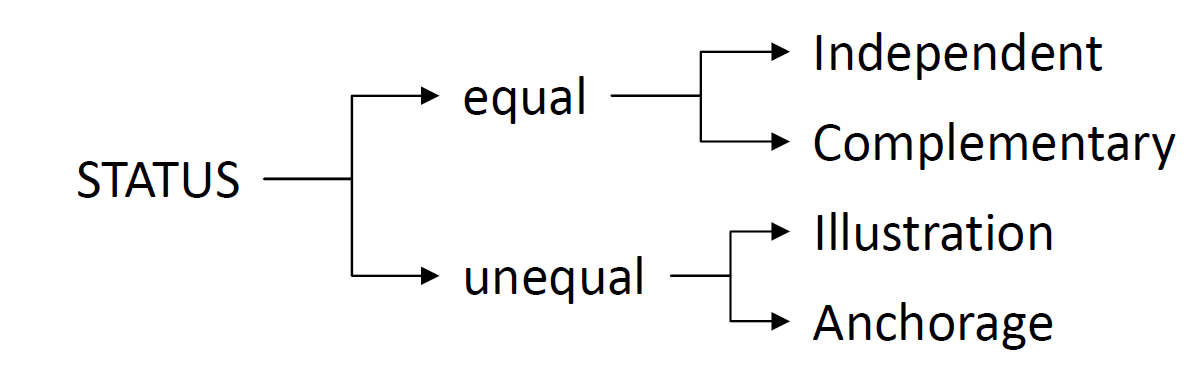}
 	\vspace{-0.3cm}
 	\caption{Part of Martinec and Salway's taxonomy that distinguishes 
 	image-text relation based on status (simplified).}
 	\label{fig:tax_martinec_salway}
 \end{figure} 
 
One direction of research in recent decades has dealt with the assignment of 
image-text pairs to distinct classes. In a pioneering work, 
Barthes~\cite{barthes1977image} discusses the respective roles and functions of 
text and images. He proposes a first taxonomy, which introduces different types 
of status relations between the modalities, denoting different hierarchic 
relations between the modalities.  In case of unequal status, the classes 
\textit{Illustration} and \textit{Anchorage} are distinguished, otherwise their 
relation is denoted as \textit{Relay}.

Martinec and Salway~\cite{Martinec2005} extend Barthes' taxonomy and further 
divide the image-text pairs of \textit{equal} rank into a 
\textit{Complementary} and \textit{Independent} class, indicating that the 
information content is either intertwined or equivalent in both modalities. 
They combine it with Halliday's~\cite{halliday2013halliday} logico-semantics 
relations, which originally have been developed to distinguish text clauses. 
Martinec and Salway revised these grammatical categories to capture the 
specific logical relationships between text and image regardless of their 
\textit{status}. McCloud~\cite{mccloud1993understanding} focuses on comic 
books, whose particular characteristic is that image and text do not share 
information by means of depicted or mentioned concepts, although they have a 
strong semantic connection. McCloud denotes this category as 
\textit{Interdependent} and argues that 'pictures and words go hand in hand to 
convey an idea that neither could convey alone'. Other authors mention the case 
of negative correlations between the mentioned/visually depicted concepts (for 
instance, N\"oth~\cite{noth1995handbook} or van 
Leeuwen~\cite{van2005introducing}), denoting them \textit{Contradiction} or 
\textit{Contrast}, respectively. Van Leeuwen states that they can be used 
intentionally, e.g., in magazine advertisements by choosing opposite colors or 
other formal features to draw attention to certain objects.

	\section{Semantic Image-Text Relations}
\label{sec:image_text_relations}
\begin{SCfigure*}[0.2][htbp]
	\includegraphics[width=0.85\textwidth]{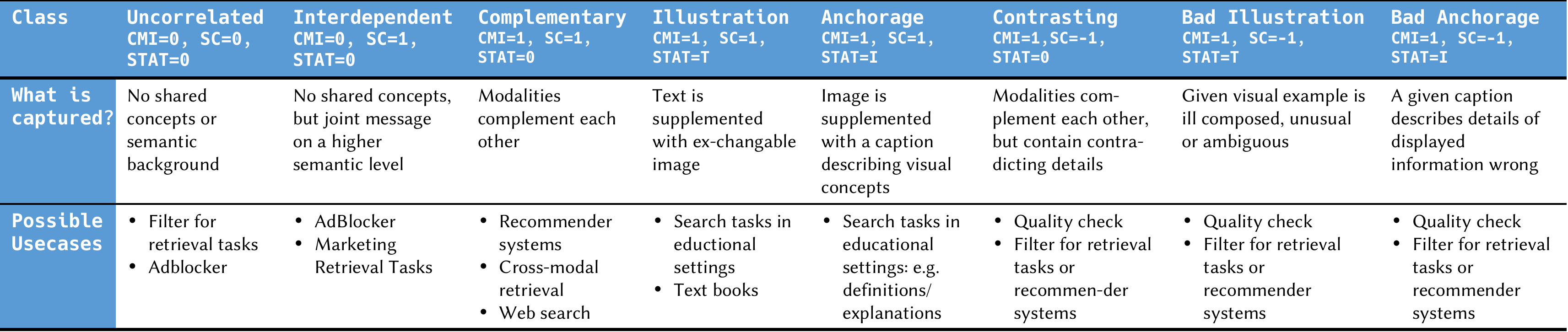}
	\caption{Overview of the proposed image-text classes and their potential 
	use cases.}
	\label{fig:class_table}
\end{SCfigure*}
The discussion of related work reveals that the complex cross-modal interplay of image and text has not been systematically modeled and investigated yet from a computer science perspective. In this section, we derive a categorization of classes of semantic image-text relations which can be used for multimedia information retrieval and web search. This categorization is based on previous work in the fields of visual communication (sciences) and information retrieval. However, one drawback of taxonomies in communication science is that their level of detail makes it sometimes difficult to assign image-text pairs to a particular class, as criticized by Bateman \cite{bateman2014text}. 

First, we evaluate the image-text classes described in communication science literature according to their usefulness for information retrieval. As a point of departure, we consider Martinec and Salway's taxonomy in the status dimension (Fig. \ref{fig:tax_martinec_salway}). This yields the classes of image-text relations of \textit{Illustration}, \textit{Anchorage}, \textit{Complementary}, and \textit{Independent}. We disregard the class \textit{Independent} since it is very uncommon that both modalities exactly describe the same information. Furthermore, we introduce the class \textit{Interdependent} as suggested by McCloud, which in contrast to \textit{Complementary} consists of image-text pairs where the intended meaning cannot be gathered from either of them exclusively. 
While a number of categorizations does not consider negative semantic correlations at all, N\"oth~\cite{noth1995handbook}, van Leeuwen~\cite{van2005introducing}, and Henning and Ewerth~\cite{henning2017estimating} consider this aspect. We believe that it is important for information retrieval tasks to consider negative correlations as well, for instance, in order to identify less useful multimodal information, mistakes etc. Consequently, we introduce the classes \textit{Contrasting}, \textit{Bad Illustration}, and \textit{Bad Anchorage}, which are the negative counterparts for \textit{Complementary}, \textit{Illustration}, and \textit{Anchorage}. Finally, we consider the case when text and image are \textit{uncorrelated}.

While one objective of our work is to derive meaningful, distinctive and comprehensible image-text classes, another contribution is their systematic characterization. For this purpose, we leverage the metrics \textit{cross-modal mutual information} (CMI) and \textit{semantic correlation} (SC)~\cite{henning2017estimating}. However, these two metrics are not sufficient to model a larger set of image-text classes. It stands out that the \textit{status} relation, originally introduced by Barthes \cite{barthes1977image}, is adopted by the majority of taxonomies established in the last four decades (e.g. \cite{Martinec2005, unsworth2007image}), implying that this relation is essential to describe an image-text pair. It portrays how two modalities can relate to one another in a hierarchical way reflecting their relative importance. Either the text supports the image (\textit{Anchorage}), or the image supports the text (\textit{Illustration}), or both modalities contribute equally to the overall meaning (e.g., \textit{Complementary}, originally denoted by Barthes as \textit{Relay}). This encourages us to extend the two-dimensional feature space of CMI and SC with the \textit{status} dimension (\textit{STAT}). In the next section, we provide some definitions for the three metrics and subsequently infer a categorization of semantic image-text classes from them. Our goal is to reformulate and clarify the interrelations between visual and textual content in order to make them applicable for multimodal indexing and retrieval. An overview of the image-text classes and their mapping to the metrics, as well as possible use cases is given in Figure \ref{fig:class_table}.

\subsection{\textbf{Metrics for image-text relations}}
\label{sec:metrics}
 
\textbf{Concepts and entities:} 
The following definitions are related to concepts and entities in images and text. Generally, plenty of concepts and entities can be found in images ranging from the main focus of interest (e.g., a person, a certain object, an event, a diagram) to barely visible or background details (e.g., a leaf of grass, a bird in the sky). 
Normally, the meaning of an image is related to the main objects in the foreground. When assessing relevant information in images, it is reasonable to regard these concepts and entities, which, however, adds a certain level of subjectivity in some cases. But most of the time the important entities can be easily determined.

\textbf{Cross-modal mutual information (CMI)}
\newline Depending on the (fraction of) mutual presence of concepts and entities in both image and text, the cross-modal mutual information ranges from $0$ (no overlap of depicted concepts) to $1$ (concepts in image and text overlap entirely).

It is important to point out that CMI ignores a deeper semantic meaning, in contrast to \textit{semantic correlation}. If, for example, a small man with a blue shirt is shown in the image, while the text talks about a tall man with a red sweater, the CMI would still be positive due to the mutual concept "man". But since the description is confusing and hinders interpretation of the multimodal information, semantic correlation (SC, see below) of this image-text pair would be negative. Image-text pairs with high CMI can be found in image captioning datasets, for instance. The images and their corresponding captions have a descriptive nature, that is they have explicit representations in both modalities. In contrast, news articles or advertisements often have a rather loose connection to their associated images by means of mutual entities or concepts. The range of cross-modal mutual information (CMI) is $[0,1]$. 

\textbf{Semantic correlation (SC)} 
\newline The (intended) meaning of image and text can range from coherent (SC=$1$), over independent (SC=$0$)  to contradictory (SC=$-1$). This refers to concepts and entities, descriptions and interpretation of symbols, metaphors, as well as to their relations to one another.

Typically, an interpretation requires contextual information, knowledge, or experience and it cannot be derived exclusively from the entities in the text and the objects depicted in the image. Possible values range from $[-1,1]$, where a negative value indicates that the co-occurrence of an image and a text disturbs the comprehension of the multimodal content. This is the case if a text refers to an object in an image and cannot be found there, or has different attributes as described in the text. An observer might notice a contradiction and ask herself "Do image and text belong together at all, or were they placed jointly by mistake?". A positive score on the contrary suggests that both modalities share a semantic context or meaning. The third possible option is that there is no semantic correlation between entities in the image and the text, then $SC = 0$. 

\textbf{Status (STAT)}
\newline Status describes the hierarchical relation between an image and text with respect to their relative importance. Either the image is "subordinate to the text" ($stat=T$), implying an exchangeable image which plays the minor role in conveying the overall message of the image-text pair, or the text is "subordinate to the image" ($stat=I$), usually characterizing text with additional information (e.g., a caption) for an image that is the center of attention. An \textit{equal status} ($stat=0$) describes the situation where image and text are equally important for the overall message.

Images which are "subordinate to text" (class \textit{Illustration}) 'elucidate' or 'realize' the text. This is the case, if a text describes a general concept and the associated image shows a concrete example of that concept. Examples for \textit{Illustrations} can be found in textbooks and encyclopedias. On the contrary, in the class \textit{Anchorage} the text is "subordinate to the image". This is the case, if the text answers the question "What can be seen in this image?". It is common that direct references to objects in the image can be found and the readers are informed what they are looking at. This type of image-text pair can be found in newspapers or scientific documents, but also in image captioning data sets. The third possible state of a \textit{status relation} is "equal", which describes an image-text pair where both modalities contribute individually to the conveyed information. Also, either part contains details that the other one does not. According to Barthes'~\cite{barthes1977image}, this class describes the situation, where the information depicted in either modality is part of a more general message and together they elucidate information on a higher level that neither could do alone. 

\subsection{\textbf{Defining classes of image-text relations}}
\label{sec:categorization}
In this section, we show how the combination of  our three metrics can be naturally mapped to distinctive image-text classes (see also Fig. \ref{fig:class_table}). For this purpose, we simplify the data value space for each dimension. The level of semantic correlation can be modeled by the interval $[-1,1]$. Henning and Ewerth ~\cite{henning2017estimating} distinguish five levels of CMI and SC. In this work, we omit these intermediate levels since the general idea of positive, negative, and uncorrelated image-text pairs is sufficient for the task of assigning image-text pairs to distinct classes. Therefore, the possible states of semantic correlation (SC) are: $sc \in \left\{-1, 0, 1\right\}$. For a similar reason, finer levels for CMI are omitted, resulting in two possible states for $cmi \in \left\{0, 1\right\}$, which correspond to \textit{no overlap} and \textit{overlap}. Possible states of status are $stat \in \left\{T, 0, I\right\}$: \textit{image subordinate to text} ($stat=T$), \textit{equal status} ($stat=0$), and \textit{text subordinate to image} ($stat=I$).

\begin{figure}[htbp]
	\centering
	\includegraphics[width=0.4\textwidth]{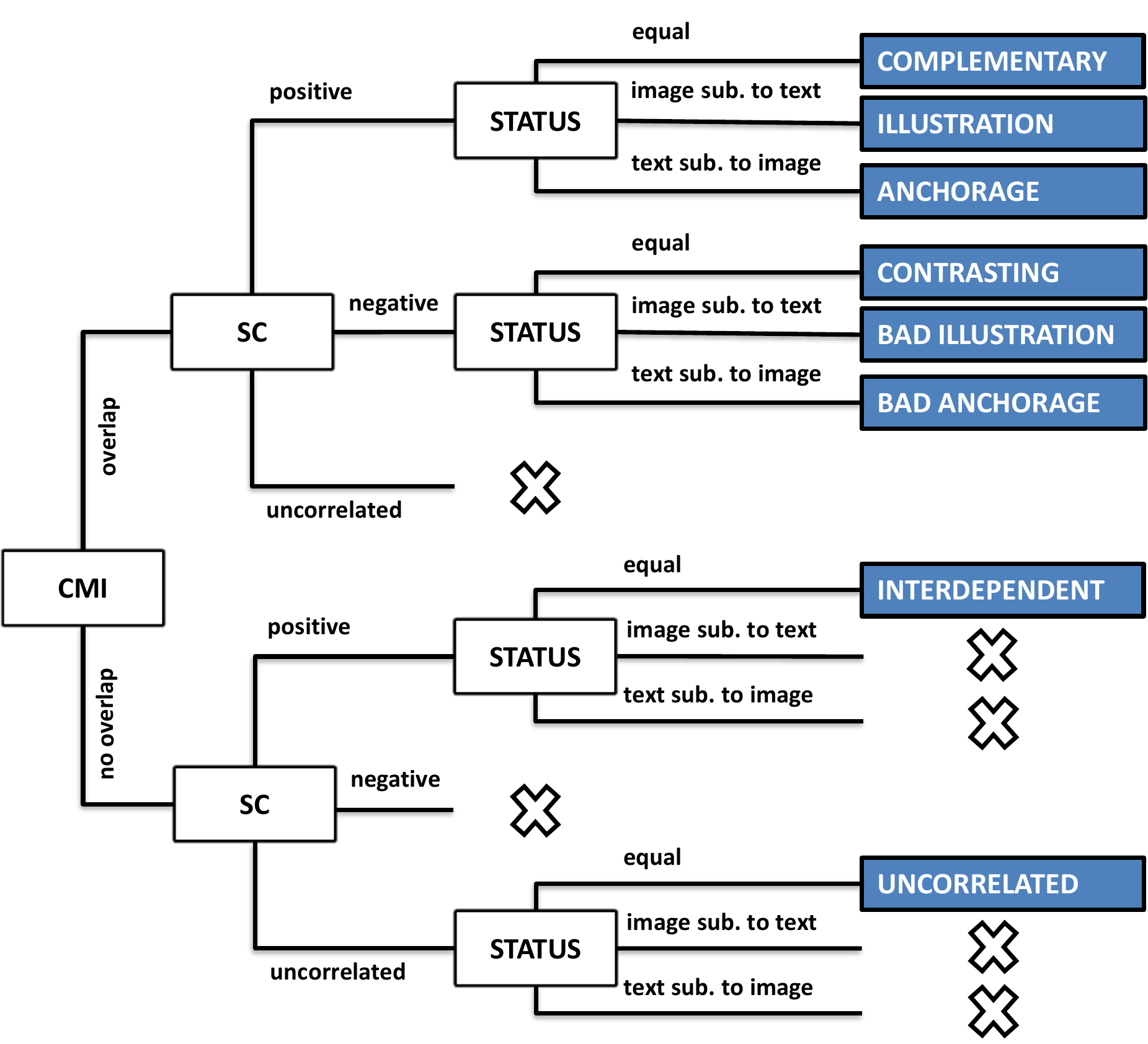}
	\vspace{-0.3cm}
	\caption{Our categorization of image-text relations. Discarded subtrees are 
	marked by an \textbf{X} for clarity. Please note that there are no 
	hierarchical relations implied.}
	\label{fig:categorization}
	\vspace{-0.2cm}
\end{figure}

If approached naively, there are $2\times3\times3=18$ possible combinations of SC, CMI and STAT. A closer inspection reveals that (only) eight of these classes match with existing taxonomies in communication sciences, confirming the coherence of our analysis. The remaining ten classes can be discarded since they cannot occur in practice or do not make sense. The reasoning behind is given after we have defined the eight classes that form the categorization.

\begin{figure}[!ht]
	\centering
	\includegraphics[width=0.4\textwidth]{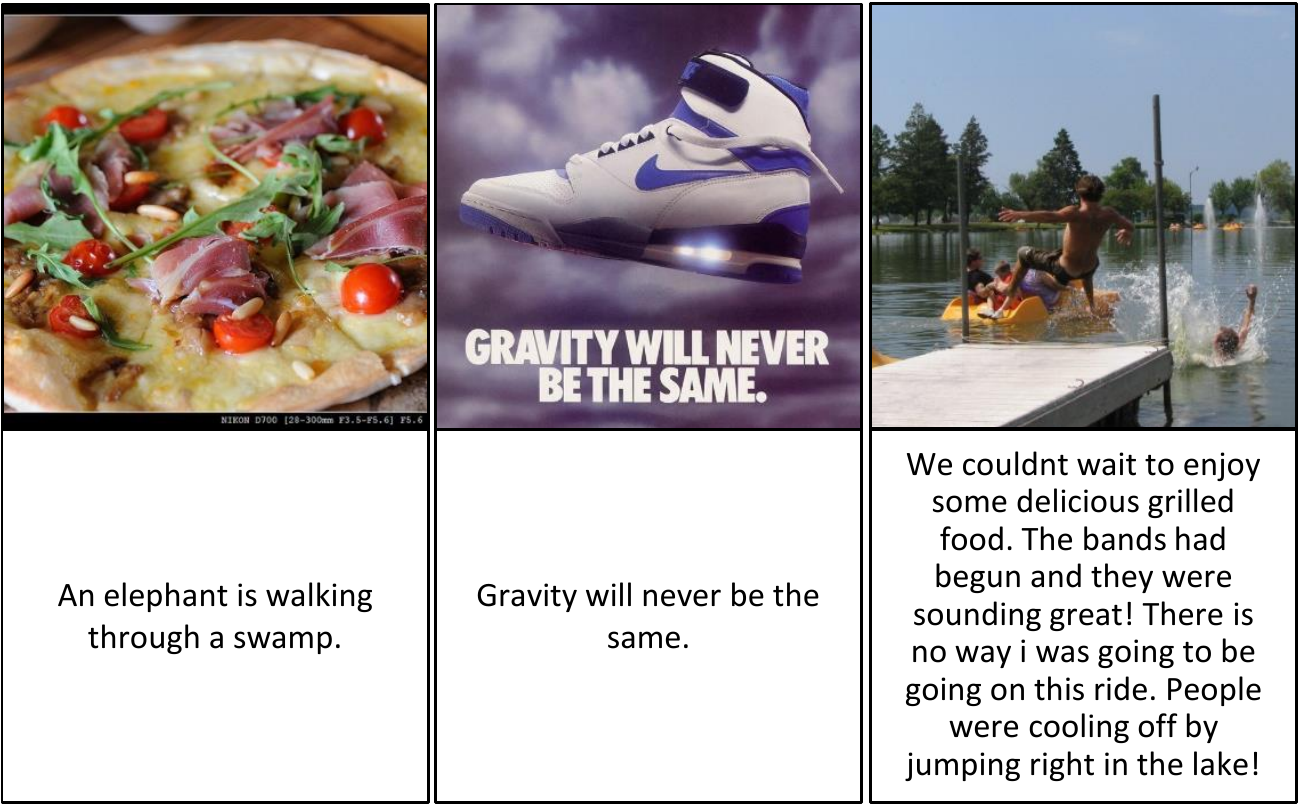}
	\caption{Examples for the \textit{Uncorrelated} (left), \textit{Interdependent} (middle) and \textit{Complementary} (right) classes. (Sources: see Section \ref{sec:train-data-augment})}
	\label{fig:example_uncorrelatedinterdependentcomplementary}
\end{figure}

\textbf{Uncorrelated ($cmi=0, sc=0, stat=0$)}\\
This class contains image-text pairs that do not belong together in an obvious way. They neither share entities and concepts nor there is an interpretation for a semantic correlation (e.g., see Fig.~\ref{fig:example_uncorrelatedinterdependentcomplementary}, left).

\textbf{Complementary ($cmi=1, sc=1, stat=0$)}\\
The class \textit{Complementary} comprises the classic interplay between visual and textual information, where both of them share information but also provide information that the other one does not. Neither of them is dependent on the other one and their status is equal. It is important to note that the amount of information is not necessarily the same in both modalities. The most significant factor is that observer is still able to understand the key information provided by either of the modalities alone (Figure~\ref{fig:example_uncorrelatedinterdependentcomplementary}, right). The definitions of the next two classes will clarify that further.

\textbf{Interdependent ($cmi=0, sc=1, stat=0$)}\\
This class includes image-text pairs that do not share entities or concepts by means of mutual information, but are related by a semantic context. As a result, their combination conveys a new meaning or interpretation which neither of the modalities could have achieved on its own. Such image-text pairs are prevalent in advertisements where companies combine eye-catching images with funny slogans supported by metaphors or puns, without actually naming their product (Figure \ref{fig:example_uncorrelatedinterdependentcomplementary}, middle). Another genre that relies heavily on these \textit{interdependent} examples are comics or graphic novels, where speech bubbles and accompanying drawings are used to tell a story. Interdependent information is also prevalent in movies and TV material in the auditory and visual modalities.

\textbf{Anchorage ($cmi=1, sc=1, stat=I$)}\\
On the contrary, the \textit{Anchorage} class is generally speaking an image description and acts as a supplement for an image. Barthes states that the role of the text in this class is to fix the interpretation of the visual information as intended by the author of the image-text pair \cite{barthes1977image}. It answers the question "What is it?" in a more or less detailed manner. This is often necessary since the possible meaning or interpretation of an image can noticeably vary and the caption is provided to pinpoint the author's intention. Therefore, an \textit{Anchorage} can be a simple image caption, but also a longer text that elucidates the hidden meaning of a painting. It is similar to \textit{Complementary}, but the main difference is that the text is subordinate to image in \textit{Anchorage}.


\textbf{Illustration ($cmi=1, sc=1, stat=T$)}\\
The class \textit{Illustration} contains image-text pairs where the visual information is subordinate to the text and has therefore a lower \textit{status}. An instance of this class could be, for example, a text that describes a general concept and the accompanying image depicts a specific example. A distinctive feature of this class is that the image is replaceable by a very different image without rendering the constellation invalid. If the text is a definition of the term "mammal", it does not matter if the image shows an elephant, a mouse, or a dolphin. Each of these examples would be valid in this scenario. In general, the text is not dependent on the image to provide the intended information.

\textbf{Contrasting ($cmi=1, sc=-1, stat=0$)}

\textbf{Bad Illustration ($cmi=1, sc=-1, stat=T$)}

\begin{figure}[htbp]
    \centering
    \includegraphics[width=0.4\textwidth]{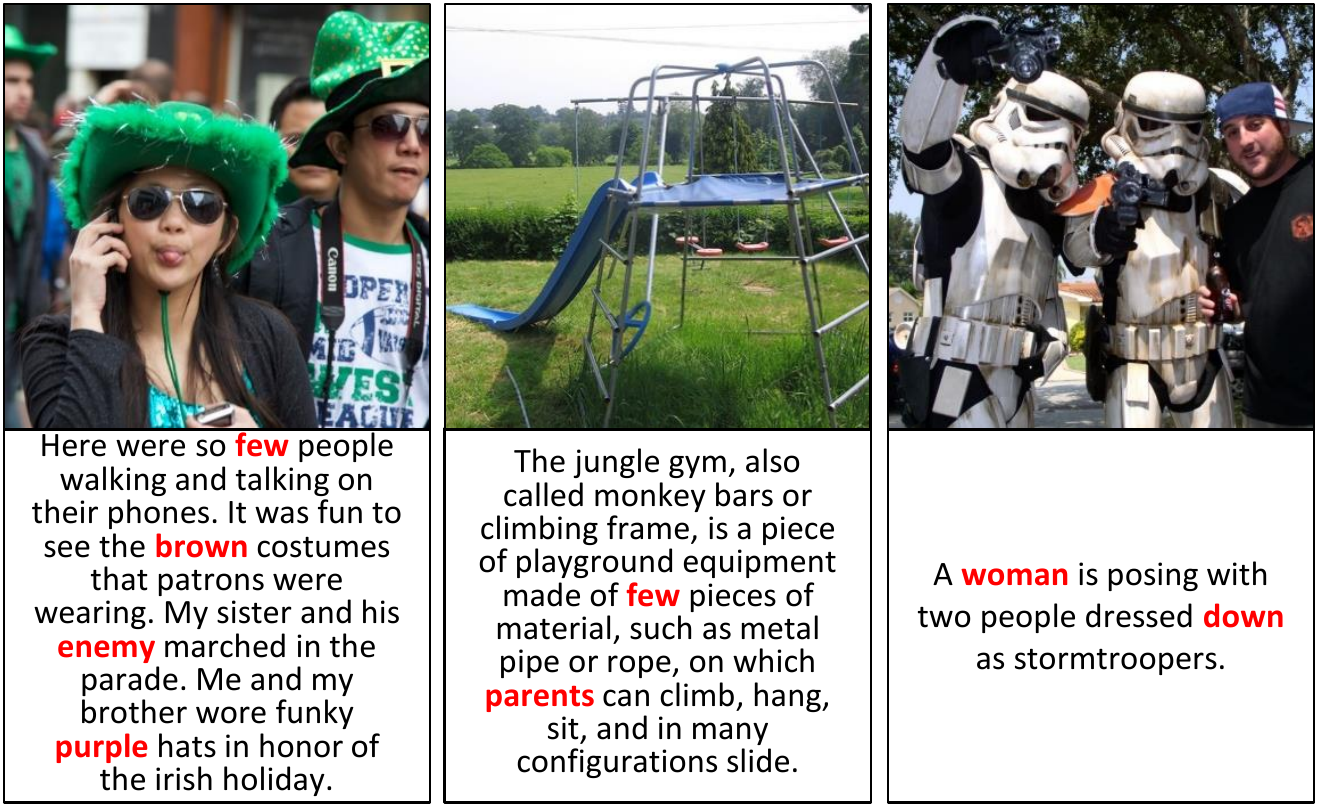}
    \vspace{-0.3cm}
    \caption{Examples for the \textit{Contrasting} (left), \textit{Bad Illustration} (middle), and \textit{Bad Anchorage} (right) classes. (Sources: see Section \ref{sec:train-data-augment})}
    \label{fig:example_contrasting_bad_illustration_bad_anchorage}
    \vspace{-0.3cm}
\end{figure}

\textbf{Bad Anchorage ($cmi=1, sc=-1, stat=I$)}\\
These three classes are the counterparts to \textit{Complementary, Illustration}, and \textit{Anchorage}: they share their primary features, but have a \textbf{negative SC} (see Fig.~\ref{fig:example_contrasting_bad_illustration_bad_anchorage}). In other words, the transfer of knowledge is impaired due to inconsistencies or contradictions when comparing image and text \cite{henning2017estimating}. In contrast to \textit{uncorrelated} image-text pairs, these classes share information and obviously they belong together in a certain way, but particular details or characteristics are contradicting. For instance, a \textit{Bad Illustration} pair could consist of a textual description of a bird, whose most prominent feature is its colorful plumage, but the bird in the image is actually a grey pigeon. This can be confusing and an observer might be unsure if he is looking at the right image. Similarly, contradicting textual counterparts exist for each of these classes. In section \ref{sec:train-data-augment}, we describe how we generate training samples for these classes.
\subsection{\textbf{Impossible image-text relations}}
The eight classes described above form the categorization as shown in Figure \ref{fig:categorization}. The following ten combinations of metrics were discarded, since they do not yield meaningful image-text pairs.\\
\textbf{Cases A: $cmi=0, sc=-1, stat=T,0,I$}\\
These three classes cannot exist: If the shared information is zero, then there is nothing that can contradict one another. As soon as a textual description relates to a visual concept in the image, there is cross-modal mutual information and $CMI > 0$.\\
\textbf{Cases B: $cmi=0, sc=0, stat=T,I$}\\
The metric combination $cmi=0, sc=0, stat=0$ describes the class \textit{Uncorrelated} of image-text pairs which are neither in contextual nor visual relation to one another. Since it is not intuitive that a text is subordinate to an uncorrelated image or vice versa, these two classes are discarded.\\
\textbf{Cases C: $cmi=0, sc=1, stat=T,I$}\\
Image-text pairs in the class \textit{Interdependent} ($cmi=0, sc=1, stat=0$) are characterized by the fact, that even though they do not share any information they still complement each other by conveying additional or new meaning. Due to the nature of this class a subordination of one modality to the other one is not plausible: Neither of the conditions for the states \textit{image subordinate to text} and \textit{text subordinate to image} is fulfilled due to lack of shared concepts and entities. Therefore, these two classes are discarded.\\
\textbf{Cases D: $cmi=1, sc=0, stat=T,0,I$}\\
As soon as there is an overlap of essential depicted concepts there has to be a minimum of semantic overlap. We consider entities as essential, if they contribute to the overall information or meaning of the image-text pair. This excludes trivial background information such as the type of hat a person wears in an audience behind a politician giving a speech. The semantic correlation can be minor, but it would still correspond to $SC=1$ according to the definition above. Therefore, the combination $cmi=1, sc=0$ and the involved possible combinations of \textit{STAT} are discarded.

 	\section{Predicting Image-Text Classes}
\label{sec:classifier}
In this section, we present our approach to automatically predict the 
introduced image-text metrics and classes. We propose a deep learning 
architecture that realizes a multimodal embedding for textual and graphical 
data. Deep neural networks achieve better results, when they are trained with a 
large amount of data. However, for the addressed task no such dataset exists. 
Crowdsourcing is an alternative to avoid the time-consuming task of manually 
annotating training data on our own, but requires significant efforts to 
maintain the quality of annotations obtained in this way. Therefore, we follow 
two strategies to create a sufficiently large training set. First, we 
automatically collect image-text pairs from different open access Web sources. 
Second, we suggest a method for training data augmentation (Section 
\ref{sec:train-data-augment}) that allows us to also generate samples for the 
image-text classes that rarely occur on the Web, for instance, \textit{Bad 
Illustration}. We suggest two classifiers, a \textbf{"classic"} approach, which 
simply outputs the most likely image-text class, as well as a cascaded approach 
based on classifiers for the three metrics. The motivation for the latter is to 
divide the problem into three easier classification tasks. Their subsequent 
\textbf{"cascaded"} execution will still lead us to the desired output of 
image-text classes according to Figure~\ref{fig:categorization}. The deep 
learning architecture is explained in section \ref{sec:cascade-deep}.

\subsection{\textbf{Training data augmentation}}
\label{sec:train-data-augment}
The objective is to acquire a large training dataset of high quality image-text 
pairs with a minimum effort in manual labor. 
On the one hand, there are classes like \textit{Complementary} or 
\textit{Anchorage} available from a multitude of sources and can therefore be 
easily crawled. Other classes like \textit{Uncorrelated} do not naturally occur 
in the Web, but can be generated with little effort. On the other hand, there 
are rare classes like \textit{Contrasting} or \textit{Bad Anchorage}. While 
they do exist and it is desirable to detect these image-text pairs as well (see 
Fig.~\ref{fig:class_table}), there is no abundant source of such examples that 
could be used to train a robust classifier.

Only few datasets are publicly available that contain images and corresponding 
textual information, which are not simply based on tags and keywords but also 
use cohesive sentences. Two examples are the image captioning dataset 
MSCOCO~\cite{linMicrosoft2014} as well as the Visual Storytelling dataset 
(VIST~\cite{huang2016visual}). A large number of examples can be easily taken 
from these datasets, namely for the classes \textit{Uncorrelated}, 
\textit{Complementary}, and \textit{Anchorage}. Specifically, the underlying 
hierarchy of MSCOCO is exploited to ensure that two randomly picked examples 
are not semantically related to one another, and then join the caption of one 
sample with the image of the other one to form \textit{Uncorrelated} samples. 
In this way, we gathered $60\,000$ \textbf{\textit{uncorrelated}} training 
samples. 

The VIST dataset has three types of captions for their five-image-stories. The 
first one "Desc-in-Isolation" resembles the generic image-caption dataset and 
can be used to generate examples for the class \textbf{\textit{Anchorage}}. 
These short descriptions are similar to MSCOCO captions, but slightly longer, 
so we decide to use them. Around $62\,000$ examples have been generated this 
way. The pairs represent this class well, since they include textual 
descriptions of the visually depicted concepts without any low-level visual 
concepts or added interpretations. More examples could have been generated 
similarly, but we have to restrict the level of class imbalance. The second 
type of VIST captions "Story-in-Sequence" is used to create 
\textbf{\textit{Complementary}} samples by concatenating the five captions of a 
story and pairing them randomly with one of the images of the same story. Using 
this procedure, we generated $33\,088$ examples. 

While there are certainly more possible constellations of 
\textit{complementary} content from a variety of sources, the various types of 
stories of this dataset give a solid basis. The same argumentation holds for 
the \textbf{\textit{Interdependent}} class. Admittedly, we had to manually 
label a set of about $1\,007$ entries of Hussain et al.'s Internet 
Advertisements data set~\cite{hussain2017automatic} to generate these 
image-text pairs. While they exhibit the right type of image-text relations, 
the accompanied slogans (in the image) are not annotated separately and optical 
character recognition does not achieve high accuracy due to ornate fonts etc. 
Furthermore, some image-text pairs had to be removed, since some slogans 
specifically mention the product name. This contradicts the condition that 
there is no overlap between depicted concepts and textual description, i.e., 
\textit{cmi}$=0$. 

The \textbf{\textit{Illustration}} class is established by combining one random 
image for each concept of the ImageNet dataset~\cite{ILSVRC15} with the summary 
of the corresponding article of the English Wikipedia, in case it exists. This 
nicely fits the nature of the class since the Wikipedia summary often provides 
a definition including a short overview of a concept. An image of the ImageNet 
class with the same name as the article should be a replaceable example image 
of that concept.

The three remaining classes \textbf{\textit{Contrasting, Bad Illustration}} and 
\textbf{\textit{Bad Anchorage}} occur rarely and are hard to detect 
automatically. 
Therefore, it is not possible to automatically crawl a sufficient amount of 
samples. 
To circumvent this problem, we suggest to transform the respective positive 
counterparts by replacing around $530$ keywords~ \cite{website} (adjectives, 
directional words, colors) by antonyms and opposites in the textual description 
of the positive examples to make them less comprehensible. For instance, "tall 
man standing in front of a green car" is transformed into a "small woman 
standing behind a red car". While this does not absolutely break the semantic 
connection between image and text it surely describes certain attributes 
incorrectly, which impairs the accurate understanding and subsequently 
justifies the label of \textit{sc}$=-1$. This strategy allows us to transform a 
substantial amount of the "positive" image-text pairs into their negative 
counterparts. Finally, for all classes we truncated the text if it exceeded 10 
sentences.

In total the dataset consists of $224\,856$ image-text pairs. Table 
\ref{tab:dist_classes} and \ref{tab:dist_labels} give an overview about the 
data distribution, first sorted by class and the second one according to the 
distribution of the three metrics, which were also used in our experiments.
\vspace{0.1cm}
\begin{table}[htbp]
	\parbox{0.4\linewidth}{
		\centering
		\begin{tabular}{|l | c |}
			\hline
			Class & \# Samples \\ \hline
			\textbf{Uncorrelated} 		& 60\,000 \\ \hline
			\textbf{Interdependent} 	& 1\,007 	\\ \hline
			\textbf{Complementary} 		& 33\,088	\\ \hline
			\textbf{Illustration} 		& 5\,447 \\ \hline
			\textbf{Anchorage} 			& 62\,637 \\ \hline
			\textbf{Contrasting}		& 31\,368	\\ \hline
			\textbf{Bad Illustration} 	& 4\,099	\\ \hline
			\textbf{Bad Anchorage} 		& 27\,210 \\ \hline
		\end{tabular}
		\centering
		\caption{Distribution of class labels in the generated dataset.}
		\label{tab:dist_classes}
	}\hfill
	\parbox{0.4\linewidth}{
		\centering
		\begin{tabular}{| l | c |}
			\hline
			Class & \# Samples \\ \hline
			\textbf{STAT T} 	& 125\,463 \\ \hline
			\textbf{STAT 0} 	& 9\,546	\\ \hline
			\textbf{STAT I} 	& 89\,847	\\ \hline
			\textbf{SC -1} 		& 62\,677 \\ \hline
			\textbf{SC 0} 		& 60\,000 \\ \hline
			\textbf{SC 1}		& 102\,179 \\ \hline
			\textbf{CMI 0} 		& 61\,007 \\ \hline
			\textbf{CMI 1} 		& 163\,849 \\ \hline
		\end{tabular}
		\centering
		\caption{Distribution of metric labels in the generated dataset.}
		\label{tab:dist_labels}
	}
	\vspace{-0.6cm}
\end{table}

\subsection{\textbf{Design of the deep classifiers}}
\label{sec:cascade-deep}
\begin{figure}[ht]
	\centering
	\includegraphics[width=0.45\textwidth]{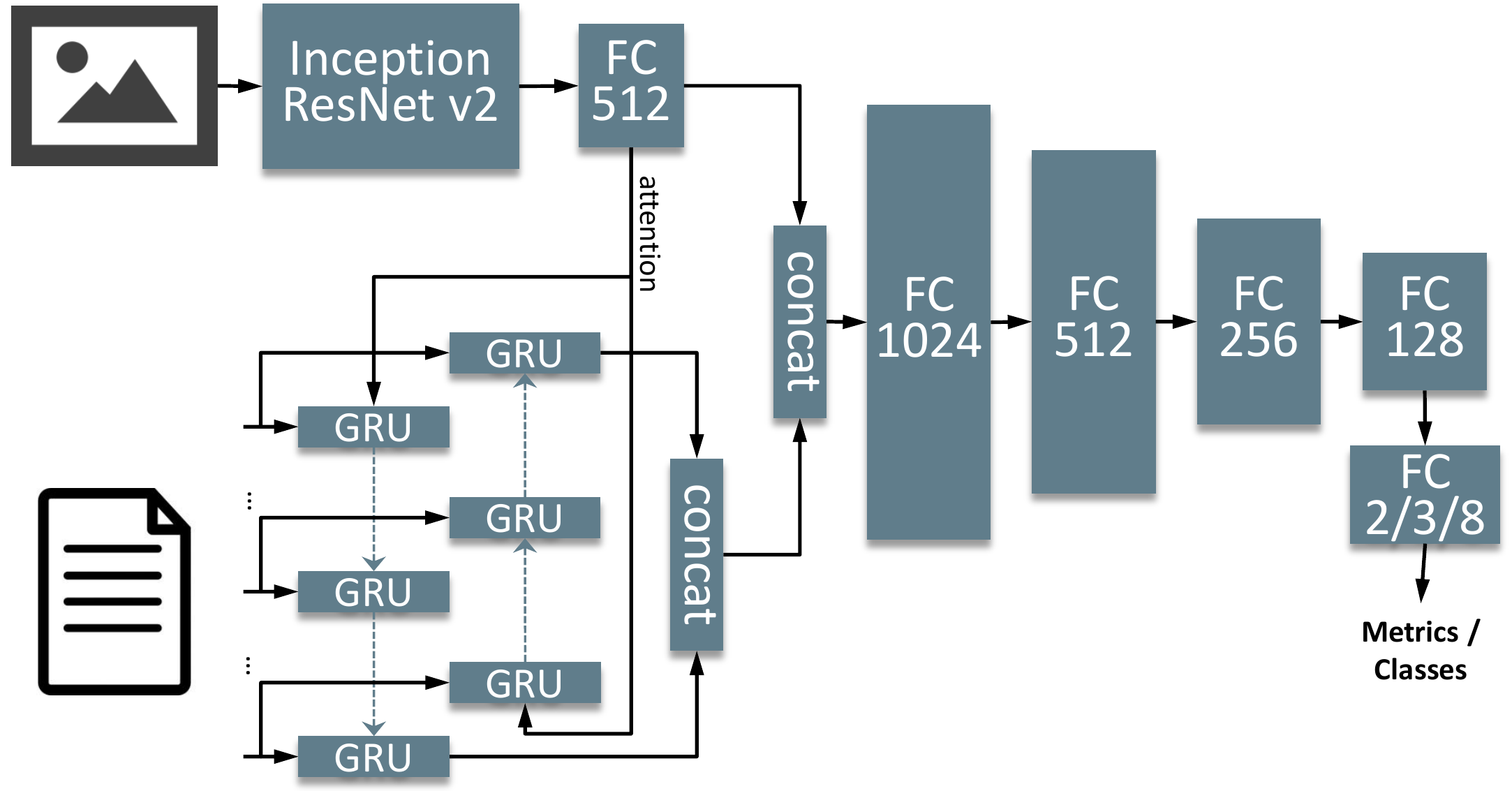}
	\caption{General structure of the deep learning system with multimodal 
	embedding. The last fully connected layer (FC) has $2$, $3$, or $8$ outputs 
	depending on whether CMI (two levels), SC/STAT (three levels), or all eight 
	image-text classes ("classic" approach) are classified.}
	\label{fig:classifier}
\end{figure}

As mentioned above, we introduce two classification approaches: "classic" and 
"cascade". The advantage of the latter is that it is easier to maintain a good 
class balance of samples, while it is also the easier classification problem. 
For instance, the classes \textit{Contrasting}, \textit{Bad Illustration}, and 
\textit{Bad Anchorage} are used to train the neural network how negative 
semantic correlation looks like. This should make the training process more 
robust against overfitting and underfitting, but naturally also increases the 
training and evaluation time by a factor of three.
Both methods follow the architecture shown in Figure \ref{fig:classifier}, but 
for "cascade" three networks have to be trained and subsequently applied to 
predict an image-text class. To encode the input image, the deep residual 
network "Inception-ResNet-v2"\,\cite{szegedy2017inception} is used, which is 
pre-trained on the dataset of the ImageNet challenge\,\cite{ILSVRC15}. To embed 
this model in our system, we remove all fully connected layers and extract the 
feature maps with an embedding size of $2048$ from the last convolutional layer.
The text is encoded by a pre-trained model of the 
word2vec~\cite{mikolov2013distributed} successor 
fastText~\cite{Bojanowski2016}, which has the remarkable ability to produce 
semantically rich feature vectors even for unknown words. This is due to its 
skip-gram technique, which does not observe words as a whole but as n-grams, 
that is a sum of word parts. 
Thus, it enables the system to recognize a word or derived phrasings despite of 
typing errors. FastText utilizes an embedding size of $300$ for each word and 
we feed them into a bidirectional GRU (gated recurrent unit) inspired by Yang 
et al.~\cite{yang2016hierarchical}, which reads the sentence(s) forwards and 
backwards before subsequently concatenating the resulting feature vectors. In 
addition, an attention mechanism is incorporated through another convolutional 
layer, which reduces the image encoding to $300$ dimensions, matching the 
dimensionality of the word representation set by fastText. In this way it is 
ensured that the neural network reads the textual information under the 
consideration of the visual features, which forces it to interpret the features 
in unison. The final text embedding has a dimension of $1024$. After 
concatenating image (to get a global feature representation from the image, we 
apply average pooling to the aforementioned last convolutional layer) and text 
features, four consecutive fully connected layers (dimensions: $1024, 512, 256, 
128$) comprise the classification layer. This layer has two outputs for 
\textit{CMI}, three outputs for \textit{SC} and \textit{STAT}, or eight outputs 
for the "classic" classifier, respectively. For the actual classification 
process in the cascade approach, the resulting three models have to be applied 
sequentially in an arbitrary order. We select the order $CMI\Rightarrow 
SC\Rightarrow STAT$, the evaluations of the three classifiers yield the final 
assignment to one of the eight image-text classes (Figure 
\ref{fig:categorization}).
	\section{Experimental Evaluation}
\label{sec:experiments}
The dataset was split into a training and test set, where the latter one was 
manually labeled to generate high quality labels. It initially contained $800$ 
image-text pairs, where for each of the eight classes 100 examples were taken 
out of the automatically crawled and augmented data. The remaining $239\,307$ 
examples were used to train the four different models (three for the "cascade" 
classifier and one for the "classic" approach) for $100\,000$ iterations each 
with the TensorFlow framework~\cite{tensorflow2015}. The \textit{Adam 
optimizer} was used with its standard learning rate and a dropout rate of $0.3$ 
for the image embedding layer and $0.4$ for the text embedding layer. Also, a 
softmax cross entropy loss was used and a batch size of $12$ on a NVIDIA Titan 
X. All images were rescaled to a size of $299\times299$ and Szegedy et 
al.'s~\cite{szegedy2015going} image preprocessing techniques were applied. This 
includes random cropping of the image as well as random brightness, saturation, 
hue and contrast distortion to avoid overfitting. In addition, we limit the 
length of the textual information to 50 words per sentence and 30 sentences per 
image-text pair. All "Inception-ResNet-v2" layers were pre-trained with the 
ILSVRC (ImageNet Large Scale Visual Recognition Competition) 
2010\,\cite{ILSVRC15} dataset to reduce the training effort. The training and 
test data are publicly available at \url{https://doi.org/10.25835/0010577}. 

\subsection{Experimental results}
\label{sec:results}

To assure highly accurate groundtruth data for our test set, we asked three 
persons of our group (one of them is a co-author) to manually annotate the 
$800$ image-text pairs. Each annotator received an instruction document 
containing short definitions of the three metrics (section \ref{sec:metrics}), 
the categorization in Figure~\ref{fig:categorization}, and one example per 
image-text class (similar to 
Figures~\ref{fig:example_uncorrelatedinterdependentcomplementary}-\ref{fig:example_contrasting_bad_illustration_bad_anchorage}).
 The inter-coder agreement has been evaluated using Krippendorff's 
alpha~\cite{krippendorff1970estimating} and yielded a value of $\alpha = 0.847$ 
(across all annotators, samples, and classes). A class label was assigned, if 
the majority of annotators agreed on it for a sample. Besides the eight 
image-text classes, the annotators could also mark a sample as \textit{Unsure} 
which denotes that an assignment was not possible. If \textit{Unsure} was the 
majority of votes, the sample was not considered for the test set. This only 
applied for two pairs, which reduced the size of the final test set to $798$.

\vspace{0.2cm}
\begin{table}[htbp]
	\resizebox{0.4\textwidth}{!}{
		\begin{tabular}{| l | c | c | c | c |}
			\hline
			Class & Uncorr. & Interdep. & Compl. & Illustration\\ \hline
			Recall & $69.2\%$ & $97.6\%$ & $83.8\%$ & $83.7\%$   \\ \hline
			Precision & $98.7\%$ & $96.3\%$ & $88.0\%$ & $80.7\%$ \\ \hline
			\#Samples & $149$ & $100$ & $106$ & $95$ \\ \hline \hline
			Class &	Anchorage & Contrasting & Bad Illu. & Bad Anch. \\ \hline
			Recall & $90.3\%$ &  $89.0\%$ & $98.6\%$ & $91.9\%$ \\ \hline
			Precision & $87.3\%$ &  $78.3\%$ & $69.0\%$ & $87.0\%$ \\ \hline
			\#Samples & $95$ & $87$ & $71$ & $95$ \\ \hline
	\end{tabular}}
	\caption{Comparison of the automatically generated labels with the 
	annotations of the three volunteers and the resulting number of samples per 
	class in the test set.}
	\label{tab:annotation}
	\vspace{-0.4cm}
\end{table}
\begin{SCtable*}
	\resizebox{0.78\textwidth}{!}{
		{\begin{tabular}{| l | c | c | c | c | c | c | c | c | c |}
				\hline
				Class & Uncorrelated & Interdep. & Compl. & 
				Illustration & 
				Anchorage & Contrasting & Bad Illust. & Bad Anch. & Sum \\ 
				\hline
				Uncorr. & \textbf{67} & 3 & 5 & 23 & 34 & 5 & 11 & 1 & 149\\ 
				\hline
				Interd. & 0 & \textbf{94} & 0 & 0 & 5 & 0 & 0 & 1 & 100\\ 
				\hline
				Compl.  & 0 & 0 & \textbf{93} & 0 & 4 & 9 & 0 & 0 & 106\\ 
				\hline
				Illus.  & 0 & 0 & 0 & \textbf{84} & 0 & 0 & 11 & 0 & 95\\ 
				\hline
				Anchor. & 2 & 2 & 0 & 2 & \textbf{83} & 0 & 0 & 6 & 95\\ 
				\hline
				Contr.  & 0 & 0 & 3 & 0 & 0 & \textbf{84} & 0 & 0 & 87\\ 
				\hline
				Bad Illus. & 0 & 0 & 0 & 2 & 0 & 0 & \textbf{69} & 0 & 71\\ 
				\hline
				Bad Anch. & 2 & 0 & 0 & 0 & 21 & 1 & 0 & \textbf{71} & 95\\ 
				\hline \hline
				Precision & $94.4\%$ & $94.9\%$ & $92.1\%$ & $75.7\%$ & 
				$56.5\%$ & $84.8\%$ & $75.8\%$ & $89.9\%$ & -\\ \hline 
				Recall & $45.0\%$ & $94.0\%$ & $87.7\%$ & $88.4\%$ & 
				$87.4\%$ & $96.5\%$ & $97.2\%$ &  $74.7\%$ & -\\ \hline
	\end{tabular}}}
	\caption{Confusion matrix for the \mbox{"classic"} classifier on the 
	testset of 798 image-text pairs. The rows show the groundtruth, while the 
	coloumns show the predicted samples.}\label{tab:confusion_matrix_classic}
	\vspace{-0.2cm}
\end{SCtable*}
Comparing the human labels with the automatically generated labels allowed us 
to evaluate the quality of the data acquisition process. Therefore we computed 
how good the automatic labels matched with the human ground truth labels (Table 
\ref{tab:annotation}). The low recall for the class \textit{Uncorrelated} 
indicates that there were uncorrelated samples in the other data sources that 
we exploited. The \textit{Bad Illustration} class has the lowest precision and 
was mostly confused with \textit{Illustration} and \textit{Uncorrelated}, that 
is the human annotators considered the automatically "augmented" samples either 
as still valid or uncorrelated.
\vspace{0.2cm}
\begin{table}[H]
	\resizebox{0.4\textwidth}{!}{
		\begin{tabular}{| l | c | c | c || c | c |}
			\hline
			Classifier                   & CMI     & SC      & STAT    & 
			Cascade         & Classic	       \\ \hline
			\textbf{Ours}                & 90.3\% & 84.6\% & 83.8\% 
			&\textbf{74.3\%} & \textbf{80.8\%}\\
			\cite{henning2017estimating} & 68.8\% & 49.6\% & -       & 
			-               & -               \\
			\hline
	\end{tabular}}
	\caption{Test set accuracy of the metric-specific classifiers and the two 
	final classifiers after $75\,000$ iterations.}
	\label{tab:results_classifiers}
	\vspace{-0.3cm}
\end{table}

\begin{figure}[htbp]
	\centering
	\includegraphics[width=0.5\textwidth]{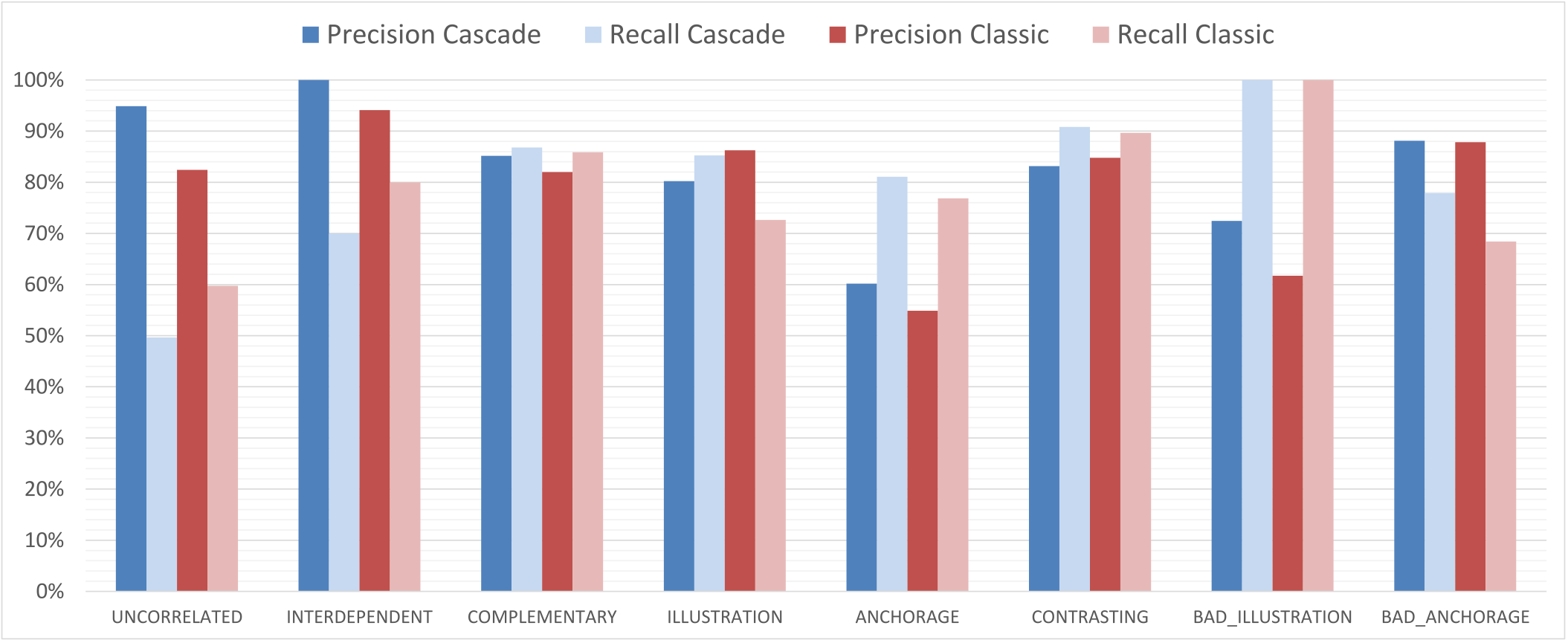}
	\vspace{-0.3cm}
	\caption{Results for both classifiers.}
	\label{fig:bar_chart}
	\vspace{-0.2cm}
\end{figure}

The (best) results for predicting image-text classes using the "classic 
approach" are presented in Table~\ref{tab:confusion_matrix_classic}. The 
overall results for our classifiers in predicting CMI, SC, STAT as well as for 
the image-text classes are presented in Table \ref{tab:results_classifiers}. 
Figure~\ref{fig:bar_chart} compares the results of the approaches "classic" and 
"cascade". The accuracy of the classifiers for CMI, SC and STAT ranges from 
83.8\% to 90.3\%, while the two classification variations for the image-text 
classes achieved an accuracy of $74.3\%$ (\emph{cascade}) and $80.8\%$ 
(\emph{classic}). We also compared our method with \cite{henning2017estimating} 
by mappingtheir  intermediate steps for CMI=$0,1,2$ to 0, CMI=3,4 to 1, and 
SC=$\pm0.5$ to $\pm1$. 
%
\vspace{0.2cm}
\subsection{Discussion of results}
As shown by Table~\ref{tab:results_classifiers} the \emph{classic} approach 
outperformed the \emph{cascade} method by about $6\%$ in terms of accuracy, 
indicating that a direct prediction of the image-text class is to be preferred 
over a combination of three separate classifiers. A reason might be that an 
overall judgment is probably more accurate than the single ones, which only 
consider one metric. This is also pleasant since an application would only need 
to train one classifiers instead of three. The class \textit{Uncorrelated} 
achieved the lowest recall indicating that both classifiers often detected a 
connection (either in the SC dimension or CMI), even though there was none. 
This might be due to the concept detector contained in InceptionResnetV2 
focusing on negligible background elements. However, the high precision 
indicates that if it was detected it was almost always correct, in particular 
for the cascade classifier. The classes with positive SC are mainly confused 
with their negative counterparts, which is understandable since the difference 
between a positive and a negative SC is often caused by a few keywords in the 
text. But the performance is still impressive when considering that positive 
and negative instances differ only in a few keywords, while image content, 
sentence length, and structure are identical.
Another interesting observation can be reported regarding the cascade approach: 
the rejection class \textit{Undefined}, which is predicted if an invalid leaf 
of the categorization (the crosses in \ref{fig:categorization}) is reached, can 
be used to judge the quality of our categorization. In total, 10 out of 18 
leaves represent such an invalid case, but only $27$ image-text pairs ($3.4\%$) 
of all test samples were assigned to it. Thus, the distinction seems to be of 
high quality. This is due to the good results of the classifiers for the 
individual metrics (Table \ref{tab:results_classifiers}).
	\section{Conclusions and Future Work}
\label{sec:conclusion}

In this paper, we have introduced a set of eight semantic image-text classes and presented an approach to automatically predict them via a deep learning system that utilizes multimodal embeddings. We have leveraged previous research in communication sciences and showed how the image-text classes can be systematically characterized by the three metrics semantic correlation, cross-modal mutual information, and the status relation. Moreover, we have outlined how to gather a large training data set for the eight classes in an (almost) automatic way by exploiting data augmentation techniques. This allowed us to train a deep learning framework with an appropriate multimodal embedding. The experimental results yielded an accuracy of $77\%$ for predicting the eight image-text classes, which demonstrates the feasibility of the proposed approach. We believe that our categorization and the automatic prediction are a solid basis to enable a multitude of possible applications in fields such as multimodal web content analysis and search, cross-modal retrieval, or search as learning. 

In the future, we will  explore further semantic relations between visual and 
textual information in order to enhance the understanding of these complex 
relationships. More diverse datasets and data generation methods should be 
included such that every possible arrangement of different information sources 
is covered, e.g., scientific documents, mainstream media etc. Finally, we will 
apply our approach to different search and retrieval scenarios. 

\begin{acks}
	Part of this work is financially supported by the Leibniz Association, 
	Germany (Leibniz Competition 2018, funding line "Collaborative Excellence", 
	project SALIENT [K68/2017]).
\end{acks}
    \newpage
	\bibliographystyle{ACM-Reference-Format}

\end{document}